\documentclass[%
 reprint,
superscriptaddress,
 amsmath,amssymb,
 aps,
 showpacs,
prb,
]{revtex4-1}

\usepackage{graphicx}
\usepackage{dcolumn}
\usepackage{bm}

\newcommand{\be}{\begin{equation}}
\newcommand{\ee}{\end{equation}}
\newcommand{\bea}{\begin{eqnarray}}
\newcommand{\eea}{\end{eqnarray}}

\begin{document}
\setlength{\unitlength}{1.2mm}
\title{Quantum Fluctuation Driven First-order Phase Transitions In Optical Lattices}
\author{Boyang Liu}
\affiliation{Beijing National Laboratory for Condensed Matter
Physics, Institute of Physics, Chinese Academy of Sciences, Beijing
100190, China}
\author{Jiangping Hu}
\affiliation{Beijing National Laboratory for Condensed Matter
Physics, Institute of Physics, Chinese Academy of Sciences, Beijing
100190, China}\affiliation{Department of Physics, Purdue University,
West Lafayette, Indiana 47907, USA}

\date{\today}
\begin{abstract}
We study quantum fluctuation driven first-order phase transitions of
a two-species bosonic system in a three-dimensional optical lattice.
Using effective potential method we find that the superfluid-Mott
insulator phase transition of one type of bosons can be changed from
second-order to first-order by the quantum fluctuations of the other
type of bosons. The study of the scaling behaviors near the quantum
critical point shows that the first-order phase transition has a
different universality from the second-order one. We also discuss
the observation of this phenomenon in the realistic cold-atom
experiments based on the \emph{in situ} density measurements.
\end{abstract}
 \maketitle

\section{Introduction}
Recently, the researches of quantum
criticality in cold-atom systems have attracted a great deal of
interest. Several schemes have been proposed to determine the
critical properties by extracting the universal scaling functions
from the atomic density profiles \cite{Zhou,Hazzard,Zhang}. The
experimental observations of quantum critical behaviors of ultracold
atoms have also been reported \cite{Donner,Zhang1}. As a clean and
highly controllable system, cold atoms can be a good play ground to
study various quantum critical behaviors.

An intriguing phenomenon near the quantum critical points (QCPs) is
the effect of quantum fluctuation driven first-order phase
transitions. The QCPs may become unstable in the appearance of
competing orders. The nature of the phase transition can be changed
from second- to first-order by the quantum fluctuations. This
phenomenon was first discussed by S. Coleman and E. Weinberg
\cite{Coleman}. They investigated a theory of a massless charged
meson coupled to the electrodynamic field using effective potential
method. Starting from a model without symmetry breaking at tree
level they found that the one-loop effective potential indicated a
new energy minimum appearing away from the origin. Independently,
Halperin, Lubensky, and Ma \cite{Halperin} discovered the same
phenomenon in the Ginzburg-Landau theory of superconductor to normal
metal transition and showed that the fluctuations of the
electromagnetic field induce a first-order transition. Quantum
fluctuation driven first-order phase transitions were also discussed
in systems with multiple coupling constants \cite{Amit, Domb}.
Recently, there have appeared more examples of the nature of the
quantum phase transition is predicted to become discontinuous as the
QCP is approached
\cite{Continentino,Ferreira,Yang,Qi,Millis,She,Diehl,Bonnes}.

In this letter we investigate the quantum fluctuation driven
first-order phase transitions of a two-species boson system in a
three dimensional optical lattice. This phenomenon has not been sufficiently
explored in condensed matter physics. With the recent progress in the researches of the quantum critical behaviors
in cold atom physics we are able to observe this phenomenon in a realistic experiment. Multi-component bosonic
systems have been studied both experimentally
\cite{Catani,Trotzky,Thalhammer,Gadway} and theoretically
\cite{Han,Buonsante,Chen,Isacsson,Li,Altman,Kuklov,Kuklov2}. Compared with the paradigmatic superfluid
to Mott insulator transition of a single component Bose gas in an optical lattice \cite{Fisher,Jaksch,Greiner,Stoferle,Spielman}, multi-component bosonic
systems have much richer phase diagrams. In our work we implement Coleman and Weinberg's effective potential method
\cite{Coleman} to calculate the quantum corrections to the classical action up to
one-loop level. We find that the superfluid-Mott insulator phase transition of one type of bosons can be driven from second-order to first-order by the quantum fluctuations of the other type. We study the scaling behaviors near the first-order phase
transition and give a feasible proposal to observe this phenomenon
in cold-atom experiments.

\section{Two-species Bose-Hubbard model}
To describe Bose-Bose
mixtures loaded into optical lattices, we consider the following
two-species Bose-Hubbard Hamiltonian, \bea
H=&&-\sum_{\alpha,<ij>}t_{ \alpha}(\hat b^\dagger_{\alpha i}\hat
b_{\alpha j}+\hat b^\dagger_{\alpha j}\hat b_{\alpha
i})-\sum_{\alpha, i}\mu_\alpha\hat n_{\alpha i}\cr&&+\sum_{\alpha,
i}\frac{U_\alpha}{2}\hat n_{\alpha i}(\hat n_{\alpha
i}-1)+U_{AB}\sum^N_{i=1}\hat n_{1i}\hat n_{2i}.\eea Here
$b^\dagger_{\alpha i}$ creates a boson of sort $\alpha=A, B$ at site
$i$. The first term in the Hamiltionian represents the hopping of
bosons of types $A$ and $B$ between the nearest-neighbor pairs of
sites $<ij>$ with hopping amplitudes $t_A$ and $t_B$. $\hat
n_{\alpha i}\equiv \hat b^\dagger_{\alpha i}\hat b_{\alpha i}$ is
the number operator of the $\alpha$ type boson at the site $i$. We
have two chemical potential $\mu_A$ and $\mu_B$ to fix the total
number of type $A$ and $B$ bosons. $U_\alpha$ and $U_{AB}$ denote
the intra- and inter-species on-site interaction strengthes.

\begin{figure}[h]
  \includegraphics[width=8.5cm]{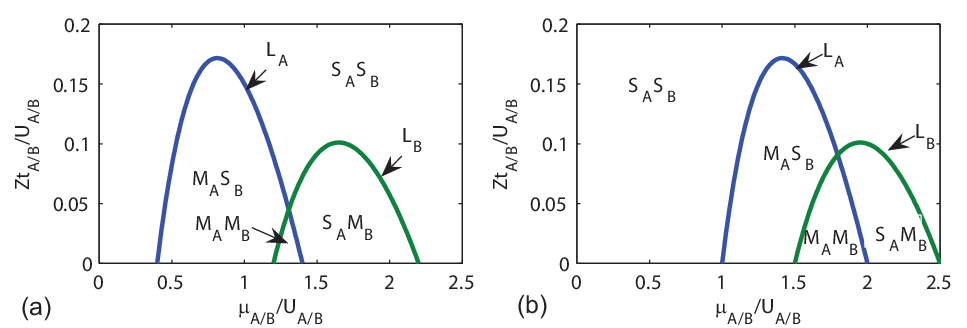}
  \caption{(Color online) The phase diagrams of the two-species Bose-Hubbard model. The axes are either $Zt_A$ versus $\mu_A$ or $Zt_B$ versus $\mu_B$ ,
all in units of $U_A$ or $U_B$, where $Z$ is the number of the
nearest neighbors around each lattice point. Curves $L_A$ and $L_B$
denote the superfluid-Mott insulator phase boundaries for species A
and B. Depending on different parameters, $L_A$ and $L_B$ may divide
the diagram into two, three, or four regions, two examples are
presented here: (a)$n^0_1=1$, $n^0_2=2$ and
$U_{AB}/U_A=U_{AB}/U_B=0.5$; (b)$n^0_1=1$, $n^0_2=2$ and
$U_{AB}/U_A=U_{AB}/U_B=0.2$. Labels ``$S_AS_B$", ``$M_AM_B$",
``$S_AM_B$" and ``$M_AS_B$" denote for superfluid phase for both
species A and B, Mott insulator phase for both species A and B,
superfluid phase for species A  Mott insulator phase for species B
and superfluid phase for species B  Mott insulator phase for species
A. }
 \label{MF}
 \end{figure}

The mean-field analysis shows that the system has three different
phases \cite{Chen}: (I) both species A and B stay in the superfluid
phases; (II) one species is in the superfluid phase and the other
one is in the Mott insulator phase; and (III) both species are in
the Mott insulator phases. Two examples of the phase diagrams are
shown in Fig. \ref{MF}.
To study the quantum fluctuation effects in the vicinity of QCPs we
may take the limit of vanishing lattice constant and finally write
down a continuum quantum field theory to describe the phase
transitions. This can be done by following a standard procedure
\cite{Sachdev}: (I) writing the partition function in the coherent
state path integral representation; (II) decoupling the hoping terms
by introducing two auxiliary fields $\varphi_1$ and $\varphi_2$
through the Hubbard-Stratanovich transformation; and (III)
integrating out the fields $b_{Ai}^\dagger$, $b_{Ai}$,
$b_{Bi}^\dagger$ and $b_{Bi}$. Then the action can be written as
\bea S=&&\int d\tau d^dx
\big\{u_1\varphi_1^\ast\partial_\tau\varphi_1+v_1|\partial_\tau\varphi_1|^2
+w_1|\nabla\varphi_1|^2\cr&&+u_2\varphi_2^\ast\partial_\tau\varphi_2+v_2|
\partial_\tau\varphi_2|^2+w_2|\nabla\varphi_2|^2+r_1|\varphi_1|^2\cr&&
+r_2|\varphi_2|^2+\frac{g_1}{2}|\varphi_1|^4+\frac{g_2}{2}|\varphi_2|^4
+g_3|\varphi_1|^2|\varphi_2|^2\big\}.\label{Ham}\eea The average of
the two Hubbard-Stratanovich field $\varphi_1$ and $\varphi_2$ are
proportional to $\langle b_A(x,\tau)\rangle$ and $\langle
b_B(x,\tau)\rangle$. Hence, they can be taken as the superfluid
order parameters. All the coefficients in Eq.(\ref{Ham}) can be
expressed in terms of the hopping amplitudes $t_\alpha$, the
chemical potentials $\mu_\alpha$ and the on-site interaction
strengths $U_\alpha$ and $U_{AB}$. \bea
r_1&&=\frac{1}{zt_A}-\frac{n^0_A+1}{\Delta_{A+}}-\frac{n^0_A}{\Delta_{A-}},\cr
r_2&&=\frac{1}{zt_B}-\frac{n^0_B+1}{\Delta_{B+}}-\frac{n^0_B}{\Delta_{A-}}
, \label{eq:r}\eea where \bea &&
\Delta_{A(B)+}=-\mu_{A(B)}+U_{A(B)}n^0_{A(B)}+U_{AB}n^0_{B(A)},\cr&&\Delta_{A(B)-}=\mu_{A(B)}-U_{A(B)}(n^0_{A(B)}-1)-U_{AB}n^0_{B(A)},\eea
which denote the particle and hole excitation energy of the species
$A(B)$. The occupation numbers $n^0_{A(B)}$ is defined as the
smallest integer larger than
$\frac{U_{AB}(U_{AB}-U_{B(A)}-2\mu_{B(A)})+2\mu_{A(B)}U_{B(A)}}{2(U_{A(B)}U_{B(A)}-U^2_{AB})}$.
The equation $r_1=0$ and $r_2=0$ generate the mean-field phase
boundaries in Fig. \ref{MF}. Furthermore, the two-species
Bose-Hubbard model obeys a $U(1)\times U(1)$ gauge symmetry, which
implies that the model is invariant under the transformation
$b_\alpha(\tau)\rightarrow b_\alpha(\tau) e^{i\theta_\alpha(\tau)}$,
$\phi_\alpha(\tau)\rightarrow
\phi_\alpha(\tau)e^{i\theta_\alpha(\tau)}$ and
$\mu_\alpha\rightarrow\mu_\alpha+i\partial_\tau\theta_\alpha(\tau)$,
where $\alpha=A,B$. This gauge invariance helps to fix the
coefficients of the first- and second-order time derivatives as
\cite{Sachdev} $u_1=-\frac{1}{t_A^2}\frac{\partial
t_A}{\partial\mu_A}$, $u_2=-\frac{1}{t_B^2}\frac{\partial
t_B}{\partial\mu_B}$, $v_1=\frac{1}{t_A^3}(\frac{\partial
t_A}{\partial\mu_A})^2-\frac{1}{2t_A^2}\frac{\partial^2
t_A}{\partial\mu_A^2}$ and $v_2=\frac{1}{t_B^3}(\frac{\partial
t_B}{\partial\mu_B})^2-\frac{1}{2t_B^2}\frac{\partial^2
t_B}{\partial\mu_B^2}$, where partial derivatives $\partial
t_{A(B)}/\partial\mu_{A(B)}$ and  $\partial^2
t_{A(B)}/\partial\mu_{A(B)}^2$ can be calculated from Eq.
(\ref{eq:r}) for fixed $r_i$. Along the mean-field phase boundaries
the parameter $u_1$ and $u_2$ can be expressed as\bea
u_1&&=\frac{z(n^0_A+1)}{\Delta_{A+}^2}-\frac{zn^0_A}{\Delta_{A-}^2},
\cr
u_2&&=\frac{z(n^0_B+1)}{\Delta_{B+}^2}-\frac{zn^0_B}{\Delta_{B-}^2}.\label{eq:u1u2}\eea
It's straight forward to see that at the tips of the insulating
lobes coefficients $u_1$ and $u_2$ vanish. For simplicity we
consider the QCPs at the tips of the insulating lobes, then he
action of Eq. (\ref{Ham}) is deduced to a relativistic theory. This
also reflects the particle-hole symmetry at the tips of the
insulating lobes. For example, we take the insulating lobes of
$n^0_A=n^0_B=1$. Using Eq. (\ref{eq:u1u2}) we obtain \be
\mu_A(\mu_B)=U_A(U_B)/(\sqrt 2+1)+U_{AB}\label{condition},\ee for
$u_1=u_2=0$. With this relations we can fine-tune the system around
the tips of the lobes. In the harmonic trap this condition locates a
shell in the cloud of gas. By varying the the optical potential
depth we will be able to change the hopping term $t_\alpha$ so that
the system can go across the phase transition point. Furthermore,
the interaction couplings can also be calculated as \bea &&
g_1=\cr&&\frac{2(n^{(0)}_A+1)^2}{\Delta_{A+}^3}+\frac{2(n^{(0)}_A)^2}{\Delta_{A-}^3}+\frac{(n^{(0)}_A+1)n^{(0)}_A}{\Delta_{A+}\Delta_{A-}}(\frac{1}{\Delta_{A+}}+\frac{1}{\Delta_{A-}}),\cr&&
g_2=\cr&&\frac{2(n^{(0)}_B+1)^2}{\Delta_{B+}^3}+\frac{2(n^{(0)}_B)^2}{\Delta_{B-}^3}+\frac{(n^{(0)}_B+1)n^{(0)}_B}{\Delta_{B+}\Delta_{B-}}(\frac{1}{\Delta_{B+}}+\frac{1}{\Delta_{B-}}),\cr&&
g_3=\frac{(n^{(0)}_A+1)(n^{(0)}_B+1)}{\Delta_{A+}\Delta_{B+}}(\frac{1}{\Delta_{A+}}+\frac{1}{\Delta_{B+}})\cr&&+
\frac{n^{(0)}_An^{(0)}_B}{\Delta_{A-}\Delta_{B-}}(\frac{1}{\Delta_{A-}}+\frac{1}{\Delta_{B-}})\cr&&+\frac{(n^{(0)}_A+1)n^{(0)}_B}{\Delta_{A+}\Delta_{B-}}(\frac{1}{\Delta_{A+}}+\frac{1}{\Delta_{B-}})
\cr&&+\frac{(n^{(0)}_B+1)n^{(0)}_A}{\Delta_{A-}\Delta_{B+}}(\frac{1}{\Delta_{A-}}+\frac{1}{\Delta_{B+}}).\label{g}\eea
In above equations we ignore the processes of two-particle or
two-hole excitations of one species since the one-particle and
one-hole excitation are dominant.

\section{The Coleman-Weinberg effective potential}

At the tips of the insulating lobes the classical potential of this
theory is posed right on the edge of the symmetry breaking, that is
$r_1=r_2=0$ in Eq. (\ref{Ham}). We wonder whether the quantum
fluctuations will break the symmetry or not. To answer this question
we implement the Weinberg and Coleman's effective potential method
\cite{Coleman} to calculate the quantum corrections to the action of
Eq. (\ref{Ham}).

The notion of the effective potential has been found to be very
useful in theories exhibiting spontaneously broken symmetry. It
allows one to calculate quantum corrections to the classical picture
of spontaneous symmetry breaking. This method is often useful in the
case with the presence of a classical external field. For instance,
a theory with a mean-field and quantum fluctuations. The effective
potential method was first developed in High energy physics
\cite{Coleman}. However, it's also widely used in condensed matter
theories.
Basically, we expand the field in terms of its mean value and
quantum fluctuations. Then we can integrate out the quantum
fluctuations to obtain an effective theory of the mean field. All
the quantum properties are incorporated in this effective theory.
The nature of the effective potential can be totally different from
the classical one. For example, the phase transition can be changed
from second order to first order
\cite{Continentino,Ferreira,Yang,Qi,Millis,She,Diehl,Bonnes}.

To obtain the effective potential we expand the fields $\varphi_1$
and $\varphi_2$ in Eq. (\ref{Ham}) in terms of their mean fields and
quantum fluctuations $\varphi_1\rightarrow\phi_1+\delta\phi_1$ and
$\varphi_2\rightarrow\phi_2+\delta\phi_2$ and keep the fluctuation
up to the second order. Then the action can be written as \bea
S[\phi_1, \phi_2]=S_0[\phi_1, \phi_2]+\frac{1}{2}\int d\tau d^3
x\delta\Phi^\dagger {\mathcal G}^{-1}\delta\Phi, \eea where
$S_0[\phi_1, \phi_2]=\int d\tau d^3
x\{|\partial_\tau\phi_1|^2+|\nabla\phi_1|^2+|\partial_\tau\phi_2|^2+|\nabla\phi_2|^2+\frac{g_1}{2}|\phi_1|^4+\frac{g_2}{2}|\phi_2|^4+g_3|\phi_1|^2|\phi_2|^2\}$.
The parameters $v_1$, $w_1$, $v_2$ and $w_2$ have been absorbed into
the coordinates. Field
$\delta\Phi^\dagger=[\delta\phi_1^\ast,\delta\phi_1,\delta\phi_2^\ast,\delta\phi_2]$
and $\delta\Phi$ is its Hermitian conjugate. The matrix $\mathcal
G^{-1}$ is
\begin{widetext}
\bea\left(\begin{array}{cccc}-\partial^2+2g_1\phi_1^\ast\phi_1+g_3\phi_2^\ast\phi_2 & g_1\phi_1\phi_1 & g_3\phi_1\phi_2^\ast & g_3\phi_1\phi_2\\
g_1\phi_1^\ast\phi_1^\ast &
-\partial^2+2g_1\phi_1^\ast\phi_1+g_3\phi_2^\ast\phi_2
& g_3\phi_1^\ast\phi_2^\ast & g_3\phi_1^\ast\phi_2 \\
g_3\phi_1^\ast\phi_2 & g_3\phi_1 \phi_2 &
-\partial^2+2g_2\phi_2^\ast\phi_2+g_3\phi_1^\ast\phi_1 & g_2
\phi_2\phi_2\\ g_3\phi_1^\ast\phi_2^\ast & g_3\phi_1\phi_2^\ast &
g_2\phi_2^\ast\phi_2^\ast &
-\partial^2+2g_2\phi_2^\ast\phi_2+g_3\phi_1^\ast\phi_1
\end{array}\right),\eea
\end{widetext} where $\partial^2=\partial_\tau^2+\nabla^2$.

After we integrate out the fluctuation fields $\delta\Phi$ the
effective potential of our action up to one-loop level can be
calculated as \bea &&V_{\rm
eff}=g_1/2(\phi^\ast_1\phi_1)^2+g_2/2(\phi^\ast_2\phi_2)^2+g_3\phi^\ast_1\phi_1\phi^\ast_2\phi_2\cr&&+\frac{1}{64\pi^2}\big\{m_1^4\ln
m^2_1+m_2^4\ln m^2_2+m_+^4\ln m^2_++m_-^4\ln m^2_-
\big\}\cr&&+B_1|\phi_1|^2+B_2|\phi_2|^2+C_1|\phi_1|^4+C_2|\phi_2|^4+C_3|\phi_1|^2|\phi_2|^2,\cr&&\label{veff}\eea
where \bea &&m_1^2=g_1|\phi_1|^2+g_3|\phi_2|^2,\cr&&
m_2^2=g_2|\phi_2|^2+g_3|\phi_1|^2,\cr&&
m^2_\pm=\frac{1}{2}\big|(3g_1+g_3)|\phi_1|^2+(3g_2+g_3)|\phi_2|^2\cr&&
\pm\sqrt{[(3g_1-g_3)|\phi_1|^2-(3g_2-g_3)|\phi_2|^2]^2+16g^2_3|\phi_1|^2|\phi_2|^2}\big|.\cr&&\eea
The terms with coefficients $B_1$, $B_2$, $C_1$, $C_2$ and $C_3$ in
Eq. (\ref{veff}) are the renormalization counterterms. They can be
fixed by imposing the renormalization conditions $\frac{\partial
V_{\rm
eff}}{\partial\phi_1^\ast\partial\phi_1}\big|_{\phi_1=0,\phi_2=0}=0$,
$\frac{\partial V_{\rm
eff}}{\partial\phi_2^\ast\partial\phi_2}\big|_{\phi_1=0,\phi_2=0}=0$,
$V_{\rm eff}(|\phi_1|=M,|\phi_2|=0)=\frac{g_1}{2}M^4$, $V_{\rm
eff}(|\phi_1|=0,|\phi_2|=M)=\frac{g_2}{2}M^4$, $V_{\rm
eff}(|\phi_1|=M,|\phi_2|=M)=(\frac{g_1}{2}+\frac{g_2}{2}+g_3)M^4$,
where $M$ is the renormalization parameter and can be chosen
arbitrarily.

\begin{figure}[h]
  \includegraphics[width=8cm]{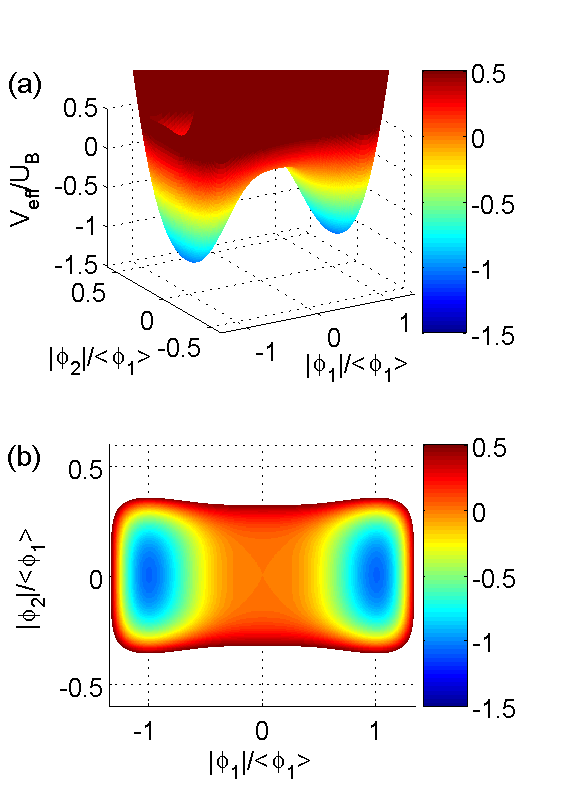}
  \caption{(Color online) The three-dimensional and contour plots of the effective potential of the two species Bose-Hubbard model.
  Coefficients $r_1=r_2=0$ for both graph (a) and (b). We use $U_B$ as the energy scale to make all the couplings dimensionless. Here we take $U_B/U_A=0.3$, then the interaction couplings are $g_2U_B^3=101.9$
  and $g_3U_B^3=26.5$.
  We take renormalization parameter $M=<\phi_1>$, where $<\phi_1>$ is the vacuum of field $\phi_1$.
  The interaction coupling $g_1$ is eliminated by the condition $\frac{\partial V_{\rm
eff}}{\partial \phi_1}|_{|\phi_1|=<\phi_1>}=0$. }
 \label{fig:veff}
 \end{figure}

The minima of the effective potential actually give the true vacuum
states with the quantum fluctuation corrections. Compared with the
classical potential where the vacuum is right at the origin, the
one-loop effective potential in Eq. (\ref{veff}) exhibits new vacua
away from the origin. This can be shown in the three-dimensional and
contour plots of the effective potential in Fig. \ref{fig:veff}.
Without loss of generality we already simplified the effective
potential by fixing the complex fields to their real directions so
that the effective potential can be easily visualized in Fig.
\ref{fig:veff}. That is, we take $\phi_1\rightarrow\phi_{1R}$ and
$\phi_2\rightarrow\phi_{2R}$. $\phi_{1R}$ and $\phi_{2R}$ are real
fields.
Here we take the parameters $g_1<g_2$ in different values then we
observe that the new vacua appear at $\phi^\ast_1\phi_1\neq0,
~\phi^\ast_2\phi_2=0$ in Fig. \ref{fig:veff} (a) and (b). Hence, the
$U(1)\times U(1)$ symmetry is spontaneously broken to $U(1)$
symmetry. At the new vacuum the field $\phi_2$ stays in the
insulator phase and field $\phi_1$ is in the superfluid phase.
 Notice that in Fig. \ref{fig:veff} we choose renormalization parameter $M=<\phi_1>$ since $M$ is arbitrary, where $<\phi_1>$ is the
 vacuum of field $\phi_1$. By setting $M=<\phi_1>$ the interaction coupling $g_1$ is eliminated through the condition $\frac{\partial V_{\rm
eff}}{\partial \phi_1}|_{|\phi_1|=<\phi_1>}=0$. Here we introduce a dimensional parameter $<\phi_1>$ and eliminate a dimensionless one $g_1$. This is called the dimensional transmutation \cite{Coleman}.

\begin{figure}[h]
\includegraphics[width=9cm]{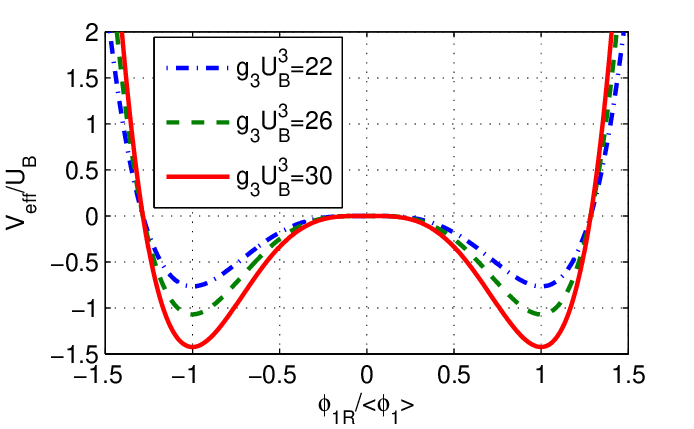}
\caption{(Color online) The effective potential along the
$\phi^\ast_2\phi_2=0$ direction with different values of $g_3$. The
parameters are $r_1=r_2=0$ and $g_2U_B^3=101.9$. $g_3$ is indicated
in the graph. We take renormalization parameter $M=<\phi_1>$. The
interaction coupling $g_1$ is eliminated by the condition
$\frac{\partial V_{\rm eff}}{\partial
\phi_1}|_{|\phi_1|=<\phi_1>}=0$.} \label{Fig:veffphi1}
\end{figure}

However, the appearance of new vacua can be an artifact since the new vacua may lie outside the range of validity of the one-loop approximation \cite{Coleman}. In order to investigate the validity of our result we
take the direction of $\phi^\ast_2\phi_2=0$ in the effective
potential to explore the vacuum. Along this direction the effective
potential can be reduced to \bea V_{\rm
eff}=&&g_1/2(\phi^\ast_1\phi_1)^2+\frac{1}{32\pi^2}g^2_3(\phi^\ast_1\phi_1)^2\ln\frac{\phi^\ast_1\phi_1}{M^2}.\label{veffphi1}
\eea The effective potential of Eq. (\ref{veffphi1}) includes a term
of $\ln\frac{\phi^\ast_1\phi_1}{M^2}$. The logarithm of a small
number is negative. Hence, the minimum arose from balancing a term of order $g_1$ against a term of order $g^2_3\ln\frac{\phi^\ast_1\phi_1}{M^2}$.
 Even though the second term formally arises in a higher order of our expansion, there is no reason why $g_1$ can not be of the same order of magnitude as
 $g_3^2$. In the realistic system the coupling constant $g_1$ and $g_3$
 can be calculated though Eq. (\ref{g}). With the condition Eq.
 (\ref{condition}) we can derive the couplings approximately as
 $g_1\sim\frac{1}{U_A^3}$ and
 $g_3\sim\frac{1}{U_AU_B}(\frac{1}{U_A}+\frac{1}{U_B})$. If we tune
 $U_A\gg U_B$ we can have $g_3^2\gg g_1$.
  Hence, our result is inside the range of validity of the one-loop approximation. The new vacuum is
illustrated in Fig. \ref{Fig:veffphi1}. As $g_3$ gets stronger the
vacuum becomes deeper.

The excitation spectrum around the new vacuum can be calculated by
expanding the effective action around the new vacuum of
$|\phi_1|^2=\rho ,~~|\phi_2|^2=0$. Let us write $
\phi_1\rightarrow\sqrt{\rho}+\delta\phi_1,
~~~~\phi_2\rightarrow\delta\phi_2$. Up to the quadratic order of the
fields $\delta\phi_1$ and $\delta\phi_2$ a straight forward
computation yields $S=\int d\tau d^3x
\big\{|\partial_\tau\delta\phi_1|^2+|\nabla\delta\phi_1|^2
+|\partial_\tau\delta\phi_2|^2+|\nabla\delta\phi_2|^2-\frac{g_3^2}{64\pi^2}\rho^2+\frac{g_3^2}{32\pi^2}\rho(\delta\phi_1^2+\delta\phi^{\ast2}_1+2\delta\phi_1^\ast\delta\phi_1)+g_3\rho\delta\phi^\ast_2\delta\phi_2\big\}.$
The diagonalization of the mass term of field $\delta\phi_1$
generates two mass eigenvalues $m_1^2=\frac{g_3^2}{16\pi^2}\rho$
or $0$. The massless excitation is the Goldstone mode, which
indicates the break down of $U(1)$ symmetry of field $\phi_1$. The
field $\delta\phi_2$ has two modes with the same mass
$m_2^2=g_3\rho$.

\section{Nature of the phase transition}
\begin{figure}[h]
\includegraphics[width=8cm]{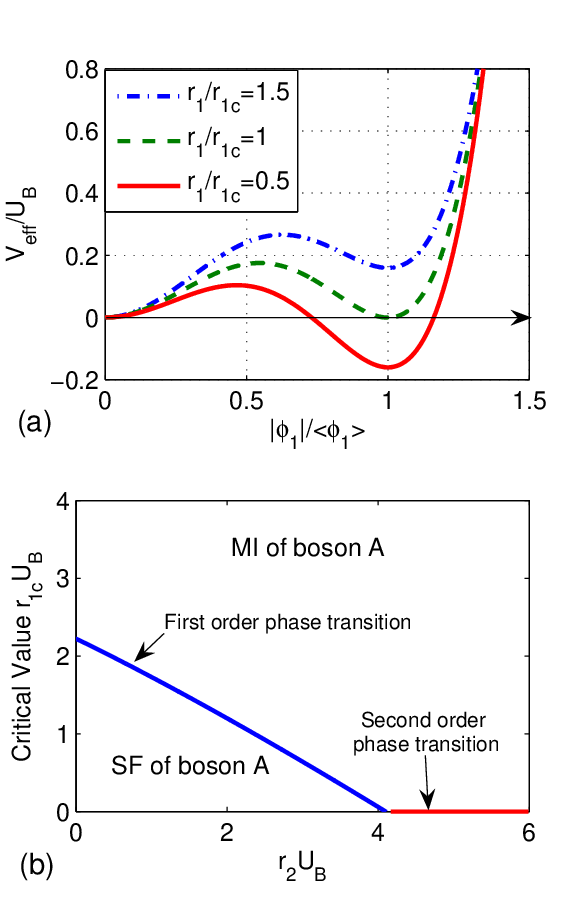}
\caption{(Color online) (a) The effective potential along the
$\phi^\ast_2\phi_2=0$ direction with different values of $r_1$,
where $r_2U_B=3$, $g_2U_B^3=101.9$, $g_3U_B^3=26.5$ and
$r_{1c}U_B\simeq0.64$. (b) The critical value $r_{1c}$ of the
first-order phase transition of type A bosons as a function of
coefficient $r_2$. }\label{fig:veffr1}
\end{figure}

We investigate the effective potential with non-zero parameter $r_1$
and $r_2$. For large enough $r_1$ and $r_2$ the vacuum of the
effective potential is at the origin. Now we vary the coefficient
$r_1$ to study how the vacuum changes. Along the direction of
$\phi^\ast_2\phi_2=0$  the effective potential is obtained as
$V_{\rm
eff}=r_1|\phi_1|^2+\frac{g_1}{2}|\phi_1|^4+\frac{(r_2+g_3|\phi_1|^2)^2}{32\pi^2}\ln(r_2+g_3|\phi_1|^2)
-\frac{r^2_2\ln r_2}{32\pi^2}-\frac{(\phi^\ast_1\phi_1)^2}{32\pi^2
M^4}(r_2+g_3M^2)^2\ln(r_2+g_3M^2).$ Here if we choose the value of
$r_1$ small enough a local minimum will appear away from the origin
as show in Fig. \ref{fig:veffr1} (a). For simplicity we take the
renormalization parameter $M^2=\langle\phi_1\rangle^2$ , where
$\langle\phi_1\rangle$ is the average value of the field $\phi_1$ at
the local minimum. Using the condition of $\frac{\partial V_{\rm
eff}}{\partial
\phi_1}|_{\phi_1^\ast\phi_1=\langle\phi_1\rangle^2}=0$ the the
effective potential can be simplified as \bea &&V_{\rm
eff}=r_1|\phi_1|^2+\frac{(r_2+g_3|\phi_1|^2)^2}{32\pi^2}\ln(r_2+g_3|\phi_1|^2)\cr&&+\frac{(\phi^\ast_1\phi_1)^2}{2\langle\phi_1\rangle^2}\Big(-r_1-\frac{g_3(r_2+g_3\langle\phi_1\rangle^2)}{16\pi^2}\ln(r_2+g_3\langle\phi_1\rangle^2)\cr&&-\frac{g_3}{32\pi^2}(r_2+g_3\langle\phi_1\rangle^2)\Big)
-\frac{r^2_2\ln r_2}{32\pi^2}.\label{veffr1}\eea As we lower the
parameter $r_1$ the vacuum of above effective potential jumps from
the origin to a new vacuum at
$\phi_1^\ast\phi_1=\langle\phi_1\rangle^2$ and
$\phi_2^\ast\phi_2=0$, where the type A bosons become superfluid and
type B bosons stay in the insulator phase. This phase transition
occurs at a finite value of $r_1$. The change of the vacuum is shown
in graph (a) of Fig. \ref{fig:veffr1}.
As $r_1$ approaches to the critical value $r_{1c}$ there is a
first-order phase transition,  where critical value of $r_1$ is \bea
&&r_{1c}=\frac{1}{16\pi^2\langle\phi_1\rangle^2}r^2_2\ln
r_2+\frac{g_3}{32\pi^2}(r_2+g_3\langle\phi_1\rangle^2)\cr&&-\frac{r_2}{16\pi^2\langle\phi_1\rangle^2}(r_2+g_3\langle\phi_1\rangle^2)\ln(r_2+g_3\langle\phi_1\rangle^2).\label{r1c}\eea
In graph (b) of Fig. \ref{fig:veffr1} we show the dependance of
$r_{1c}$ on the parameter $r_2$. As $r_2$ gets larger the critical
value $r_{1c}$ becomes smaller and even goes to zero, where the
second-order phase transition will take place. That is, if the field
$\phi_2$ is deeply in the insulator phase the first-order phase
transition of $\phi_1$ can not be induced. This quantum fluctuation
driven first-order phase transition can only happen near the QCP
with the appearance of competing orders.

At a first-order phase transition certain physical quantities, such
as the order parameter and the energy density, have a discontinuous
behavior and the correlation lengths remain generally finite. Hence,
there is no true critical behavior. However, it turns out to be
useful to develop a scaling approach for these transitions
\cite{Nienhuis,Fisher1} with scaling exponents such as $\beta=0$,
$\alpha=\gamma=1$, $\nu=1/(d+z)$ and $\delta=\infty$. In our case
the effective potential at the metastable minimum
$\phi^\ast_1\phi_1=\langle\phi_1\rangle^2$ can be written as $V_{\rm
eff}(\langle\phi_1\rangle)=1/2(r_1-r_{1c})\langle\phi_1\rangle^2$.
Introducing a parameter $\delta=r_1-r_{1c}$ which measures the
distance to the critical value $r_{1c}$, we have $V_{eff}\propto
|\delta|^{2-\alpha}$. We can identify that $\alpha=1$, which
reflects the nature of the phase transition is first order.

\begin{figure}[h]
\includegraphics[width=9cm]{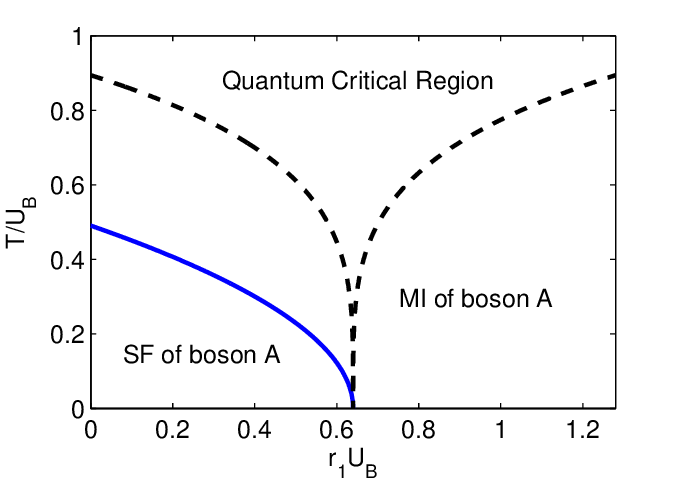}
\caption{(Color online) The phase diagram of type A bosons at finite
temperature. The parameters are set as $r_2U_B=3$, $g_2U_B^3=101.9$,
and $g_3U_B^3=26.5$. $T_c$ is the critical line of the first-order
phase transition of boson A. $T_X$ is the crossover
line.}\label{Fig:Tc}
\end{figure}

The finite temperature case can be studied through replacing the
frequency integrations in the calculation of the effective potential
by sums over the Matsubara frequencies. With high temperature
approximation $T\gg r_{1c}$ the effective potential is written as $
V_{ \rm eff}=V_{\rm
eff}(T=0)-\frac{2\pi^2}{45}T^4+(r_1+r_2+(2g_1+g_3)\phi^\ast_1\phi_1)\frac{T^2}{12},$
where $V_{\rm eff}(T=0)$ is the effective potential in
Eq.(\ref{veffr1}) and we take $k_B=1$. The first-order phase
transition at finite temperature occurs at
$r_1+\frac{T^2}{12}(2g_1+g_3)=r_{1c}$, where $r_{1c}$ is the
critical value in Eq. (\ref{r1c}). Then the critical temperature of
the first-order phase transition is \bea
T_c=\sqrt{\frac{12(r_{1c}-r_1)}{2g_1+g_3}}.\eea Furthermore, at high
temperature the effective potential at the metastable minimum can be
cast in a scaling form \bea && V_{\rm
eff}=\frac{1}{2}|\delta|^{2-\alpha}\langle\phi_1\rangle^2(1-\frac{4\pi^2T^4}{45|\delta|\langle\phi_1\rangle^2})=|\delta|^{2-\alpha}F\big[\frac{T}{T_X}\big],\cr&&\eea
where the crossover line is
$T_X=|\delta|^{z\nu}=|\delta|^{\frac{1}{4}}$. We can identify that
$\nu=1/4$ with $z=1$ in our case. This satisfies the hyperscaling
relation $2-\alpha=\nu(d+z)$ and the universality of first-order
phase transition, where $\nu=\frac{1}{d+z}$ \cite{Fisher1}. Finite
temperature phase diagram is shown in Fig. \ref{Fig:Tc}.

\section{Experimental proposals}
The study of quantum criticality in cold-atom systems is based on
\emph{in situ} density measurements
\cite{Zhou,Hazzard,Zhang,Zhang1}. General arguments show that the
observables obey universal scaling relations near the QCPs . The
density can be cast as $n(\mu, T)-n_r(\mu,
T)=T^{\frac{d}{z}+1-\frac{1}{z\nu}}G(\frac{\mu-\mu_c}{T^{1/z\nu}})$,
where $\mu_c$ is the critical value of the chemical potential, $n_r$
is the regular part of the density and $G(x)$ is a universal
function describing the singular part of the density. Following the
scheme developed by Q. Zhou and T.-L. Ho \cite{Zhou} we can plot the
``scaled density" $A(\mu, T)\equiv
T^{-\frac{d}{z}-1+\frac{1}{z\nu}}(n(\mu, T)-n_r)$ versus
$(\mu-\mu_c)/T^{\frac{1}{\nu z}}$. The scaled density curves for all
temperatures will collapse into a single curve. Here it's important
to notice that our calculation of $\nu=\frac{1}{4}$ is with respect
to the argument $\delta=r_1-r_{1c}$. However, in the realistic
cold-atom experiments we use $\mu_1-\mu_{1c}$ to measure the
distance to the QCP. Hence, a critical exponent $\tilde\nu$ with
respect to the argument $\mu_1-\mu_{1c}$ should be obtained. As we
approach the tip of the insulator lob by varying the chemical
potential we have \cite{Fisher} $\delta=r_1-r_{1c}\sim
(\mu_1-\mu_{1c})^2$
 . A straightforward calculation yields
$\tilde\nu=2\nu=\frac{1}{2}$. Then the scaled density will be in
form of $A(\mu, T)=T^{-2}(n(\mu, T)-n_r)$ near the first-order QCP,
where we have $z=1$, $d=3$ and $\tilde\nu=\frac{1}{2}$. In order to
distinguish this case from the second-order phase transition we also
calculate the scaled density near the second-order QCP, which
belongs to the three-dimensional XY universality class with critical
exponents $z=1$ and $\tilde \nu=1$ \cite{Fisher}. Then the scaled
density is $A(\mu, T)=T^{-3}(n(\mu, T)-n_r)$. By testing which form
the measured scaled density obeys we can determine whether the phase
transition is in first or second order.

\section{Conclusions}
In summary, we have investigated the quantum
fluctuation effects in two-species bosons in a three-dimensional
optical lattice. We find that nature of the superfluid-Mott
insulator phase transition of one type of bosons can be changed from
second-order to first-order by the quantum fluctuations of the other
type of bosons. The scaling behavior of this first-order phase
transition was studied and the critical exponents were calculated.
Finally, we discussed the observation of this phenomenon in a
realistic cold-atom experiment.

\section{Acknowledgements}It's a pleasure to thank Hui Zhai, Tin-Lun Ho,
Xiao-Lu Yu and Zhenhua Yu for useful discussions. This work is
supported by the NKBRSFC under grants Nos. 2012CB821400 and NSFC
under grants Nos. 1190024.

\end{document}